 \documentclass{ws-procs9x6}

\begin{document}

\title{Excited Baryons and Chiral Symmetry Breaking of QCD}

\author{Frank X. Lee}

\address{Center for Nuclear Studies,
        George Washington University, Washington, DC 20052, USA
and \\
Jefferson Lab, 12000 Jefferson Avenue, Newport News, VA 23606, USA}

\maketitle

\abstracts{
N* masses in the spin-1/2 and spin-3/2 sectors are computed
using two non-perturbative methods: lattice QCD and QCD sum rules.
States with both positive and
negative parity are isolated via parity projection methods.
The basic pattern of the mass splittings is consistent with experiments.
The mass splitting within the same parity pair is directly linked 
to the chiral symmetry breaking QCD.}

\section{Introduction}
There is increasing experimental information on the baryon spectrum
from JLab and other accelerators (as tabulated in the
particle data group~\cite{pdg00}),
and the associated desire 
to understand it from first principles.
The rich structure of the excited baryon spectrum
provides a fertile ground for exploring how the internal 
degrees of freedom in the nucleon are excited 
and how QCD works in a wider context.
One outstanding example is the parity splitting pattern
in the low-lying N* spectrum.
The splittings must be some manifestation of spontaneous chiral symmetry breaking 
of QCD because without it, QCD predicts parity doubling in the baryon spectrum.

\section{Lattice QCD}
Lattice QCD plays an important role in understanding the N* spectrum.
One can systematically study the spectrum sector by sector, with the ability 
to dial the quark masses, and dissect the degrees of freedom most responsible.
Given that state-of-the-art lattice QCD simulations have produced a ground-state
spectrum that is very close to the observed values~\cite{cppacs00},
it is important to extend the success beyond the ground states.
There exist already a number of lattice studies of the
N* spectrum~\cite{derek95,lee99,lee00,sas00,dgr00,dgr01},
focusing mostly on the spin-1/2 sector.
All established a clear splitting from the ground state.
Here, we focus on calculating
the excited baryon states in the spin-3/2 sector.
We consider the following interpolating field with the quantum numbers 
$I(J^P)={1\over 2}\left({3\over 2}^+\right)$,
\begin{equation}
\chi_\mu^N = \epsilon^{abc}( u_a^{T}C\gamma_5\gamma_\mu d_b) \gamma_5 u_c.
\end{equation}
The interpolating fields for other members of the octet can be found
by appropriate substitutions of quark fields.
Despite having an explicit parity by construction, these interpolating
fields couple to both positive and negative parity states.
A parity projection is needed to separate the two.
In the large Euclidean time limit, 
the correlator with Dirichlet boundary condition in the time direction
and zero spatial momentum becomes
\begin{equation}
G_{\mu\nu}(t) = \sum_{\bf x} <0|\chi_\mu(x)\,\bar{\chi_\nu(0)}|0>
\end{equation}
\begin{equation}
= f_{\mu\nu}\left[\lambda_+^2 {\gamma_4 +1 \over 2} e^{- M_+ \, t}
+ \lambda_-^2 {-\gamma_4 +1 \over 2} e^{- M_- \, t} \right]
\end{equation}
where $f_{\mu\nu}$ is a function common to both terms.
The relative sign in front of $\gamma_4$ provides the solution: by taking the
trace of $G_{\mu\nu}(t)$ with $(1\pm\gamma_4)/4$, one can isolate 
$M_+$ and $M_-$, respectively. 

It is well-known that a spin-3/2 interpolating field couples to both 
spin-3/2 and spin-1/2 states. 
A spin projection can be used to isolate the individual contributions 
in the correlation function $G_{\mu \nu}$~\cite{lee01}.
Numerical test of spin projection in the spin-3/2+ channel 
reveals two different
exponentials in $G(t)$  from the spin-3/2+ and spin-1/2+ parts, with the spin-3/2+
state heavier than the spin-1/2+ one, which is consistent with experiment.
One would mistake the dominant spin-1/2+ state for 
the spin-3/2+ state without spin projection.

\begin{figure}
\centerline{\psfig{file=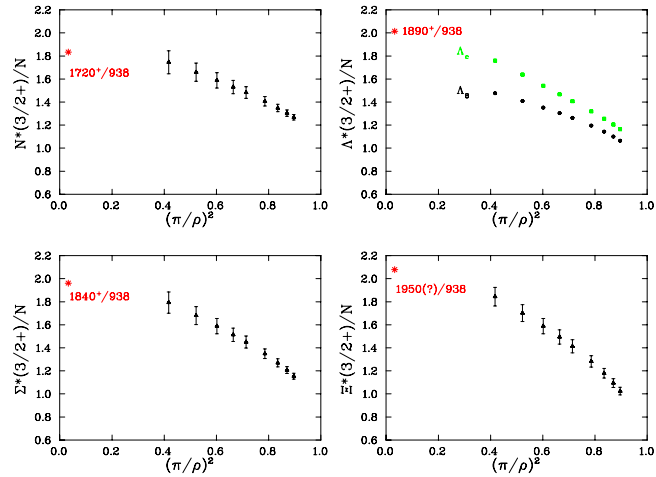,width=8.5cm,angle=90}}
\caption{Mass ratios for the $3/2+$ N* states as compared  
to experimental values where available.}
\label{Ratio4_nstar32_pos}
\end{figure}

\begin{figure}
\centerline{\psfig{file=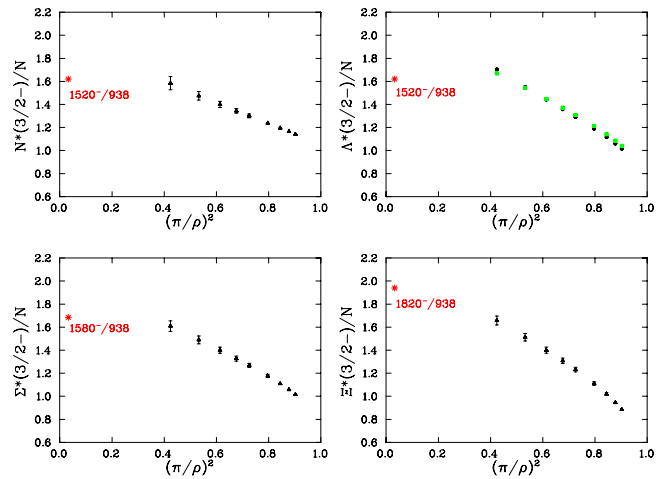,width=8.5cm,angle=90}}
\caption{Similar to Figure~\ref{Ratio4_nstar32_pos},
but for the $3/2-$ N* states.}
\label{Ratio4_nstar32_neg}
\end{figure}

Figure~\ref{Ratio4_nstar32_pos} presents preliminary results for mass
ratios extracted from the correlation functions for the $3/2^+$ N*
states to the nucleon ground state as a function of $(\pi/\rho)^2$.
Mass ratios have minimal dependence on the uncertainties in
determining the scale and the quark masses, so that a more reliable
comparison with experiment can be made.  These ratios appear headed in
the right direction compared to experiment where available, but more
study is needed to address the systematics.
Figure~\ref{Ratio4_nstar32_neg} shows the similar plots for the $3/2-$
N* states.

\section{QCD sum rules}
The QCD Sum Rule method~\cite{SVZ79} is a time-honored method
that has proven useful in revealing a connection between
hadron phenomenology and the non-perturbative nature of the QCD vacuum via
only a few parameters (the vacuum condensates).
It has been successfully applied to a variety of observables in
hadron phenomenology, providing valuable insights from a
unique, QCD-based perspective, and continues an active field.
The method is analytical (no path integrals!), is physically transparent 
(one can trace back to the QCD operators responsible),
and has minimal model dependence.
The accuracy of the approach is limited due to limitations inherent
in the operator-product-expansion (OPE), but well understood.

One progress in this area is the use of Monte Carlo-based analysis to 
explore the predictive ability of the method for 
N* properties~\cite{Derek96,Lee98,Lee01}.
The idea is to probe the entire QCD parameter space and 
map the error distribution on the OPE side to the phenomenological side.
It is found that some QCD sum rules are truly predictive for N* masses, 
while others are marginal.

\begin{figure}
\centerline{\psfig{file=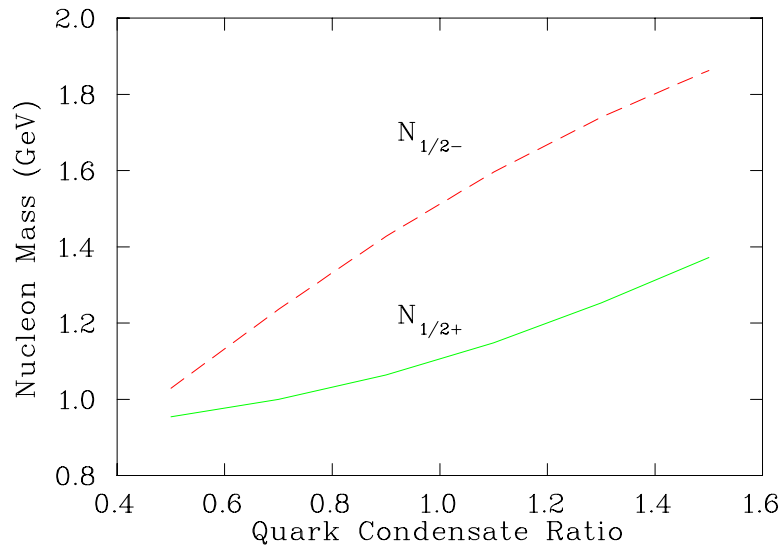,width=4.0in,angle=90}}
\caption{Mass splitting between $N_{{1\over2}-}$ and
$N_{{1\over2}+}$ as a function of the quark condensate.
The ratio is relative to the standard value of
$a=-(2\pi)^2\, \langle\bar{q}q\rangle =0.52 \mbox{ GeV}^3$.}
\label{split}
\end{figure}

Another progress is that 
a parity separation similar to that in lattice QCD can be performed 
in the QCD sum rule approach, resulting in the so-called parity-projected 
sum rules~\cite{Jido96,Jido97,Lee02} which has the general structure
\begin{equation}
{A}(M,w_+)\,+\,{B}(M,w_+)\, =\,
\tilde{\lambda}_+^2 \exp\left(-\frac{m_+^2}{M^2}\right),
\label{F_sum_rule1}
\end{equation}
\begin{equation}
{A}(M,w_-)\,-\,{B}(M,w_-)\, =\,
\tilde{\lambda}_-^2 \exp\left(-\frac{m_-^2}{M^2}\right),
\label{F_sum_rule2}
\end{equation}
where $M$ is the Borel mass parameter, 
$(m_B,\tilde{\lambda}^2,w)$ are the phenomenological parameters (mass, coupling, threshold).
The term $B$ controls the mass splitting: if $B=0$, then $m_+=m_-$.
Term $B$ involves only dimension-odd condensates, such as the quark condensate
$\langle\overline{q}q\rangle$ and the mixed condensate 
$\langle\overline{q}g\sigma\cdot Gq\rangle$.
So a direct link is established between the mass splitting of parity pairs and 
dynamical chiral symmetry breaking of QCD.
Fig.~\ref{split} shows a numerical confirmation in the case of nucleon.
As $\langle\overline{q}q\rangle$ is decreased,
both masses decrease, but with a different rate.
$N_{{1\over2}-}$ falls faster than $N_{{1\over2}+}$.
As a result, the mass splitting decreases.
In the limit that chiral-symmetry is restored ($\langle\overline{q}q\rangle$=0),
it is expected that $N_{{1\over2}+}$ and $N_{{1\over2}-}$ become degenerate.
%

In conclusion, we can compute the baryon spectrum
in the spin-1/2 and spin-3/2 sectors
for all particle channels using two methods: lattice QCD and QCD sum rules.
Parity projection further reveals that the mass splitting within the 
same baryon pair is directly controlled by spontaneous chiral symmetry breaking of QCD.

\section*{Acknowledgments}
This work is supported in part by U.S. Department of Energy 
under grant DE-FG02-95ER40907.


\end{document}